\documentclass{PoS}
\usepackage[utf8]{inputenc}
\usepackage[T1]{fontenc}
\usepackage{graphicx}
\usepackage{subcaption}
\usepackage{cite}
\usepackage{footnote}
\makesavenoteenv{tabular}

\newcommand{\be}{\begin{equation}}
\newcommand{\ee}{\end{equation}}

\title{
\begin{flushright}\begin{small}
 {\it CP3-Origins-2018-046 DNRF90\\} \vspace{1.0cm} 
  \end{small}
\end{flushright}
Hadronic spectrum calculations in the quark-gluon plasma
} 

\ShortTitle{Hadronic spectrum calculations in the quark-gluon plasma}

\author{\speaker{Jonas Glesaaen}, Gert Aarts, Chris Allton, Simon Hands, \\
    Department of Physics, College of Science, Swansea University, Swansea SA2 8PP, United Kingdom\\
     E-mail: \email{\{jonas.glesaaen,g.aarts,c.allton,s.hands\}@swansea.ac.uk}}

\author{Benjamin Jäger\\
        CP3-Origins \& Danish IAS, Department of Mathematics and
Computer Science, University of Southern Denmark, Campusvej 55, 5230 Odense M,
Denmark\\
E-mail: \email{jaeger@cp3.sdu.dk}}

\author{Jonivar Skullerud\\
        Department of Theoretical Physics, National University of Ireland Maynooth, County Kildare, Ireland\\
        E-mail: \email{jonivar@thphys.nuim.ie}}

\abstract{A status report on FASTSUM's programme of computing spectral quantities in thermal QCD, using anisotropic lattice simulations with $N_f=2+1$ flavours of Wilson fermions, is given. We provide in particular some details of the next generation of ensembles, which is currently being finalised, and give preliminary results for susceptibilities and baryonic correlators on those ensembles.

}

\FullConference{
The 36th Annual International Symposium on Lattice Field Theory - LATTICE2018\\
22-28 July, 2018\\
Michigan State University, East Lansing, Michigan, USA.
}

\begin{document}

\section{Introduction}

The goal of the FASTSUM collaboration \cite{fastsum} is to study spectral properties in thermal QCD using Wilson-type fermions on anisotropic lattices, with $a_\tau/a_s\ll 1$. We employ a fixed-scale approach, in which the temperature is varied by changing $N_\tau$, using the standard relation $T=1/(a_\tau N_\tau)$. 
This is complementary to the approach usually taken for studies of QCD thermodynamics \cite{Borsanyi:2010cj,Bazavov:2014pvz}, which utilise staggered-type fermions on isotropic lattices and vary the temperature by changing the lattice spacing.
The benefit of the anisotropic, fixed-scale approach is that it is easy to compare ensembles at different temperatures, without the need to change the bare parameters. A disadvantage is that it is not easy to reach the continuum limit, since an expensive tuning of the bare parameters at $T=0$ (gauge and fermion anisotropies, quark masses) is required for each value of the lattice spacing. However, once this has been achieved, the extension to $T>0$ is straightforward.
In previous work FASTSUM has considered bottomonium \cite{Aarts:2010ek,Aarts:2011sm,Aarts:2013kaa,Aarts:2014cda}, transport (conductivity, charge diffusion) \cite{Amato:2013naa,Aarts:2014nba},   positive- and negative-parity light baryons \cite{Aarts:2015mma,Aarts:2017rrl} and hyperons \cite{Aarts:hype}, and hidden and open charm \cite{Kelly:2018hsi}.
Here we give an update of the next generation of ensembles and present preliminary results for susceptibilities and baryons on those.

\section{Finite-temperature ensembles}

We follow the HadSpec collaboration \cite{Edwards:2008ja,Lin:2008pr,Wilson:2015dqa,Cheung:2016bym} and use a Symanzik-improved anisotropic gauge action with tree-level mean-field coefficients and a mean-field--improved Wilson clover fermion action with stout-smeared links. 
The so-called {\em Generation 2} ensembles were generated with a pion mass of $m_\pi=384(4)$ MeV, and a physical strange quark \cite{Edwards:2008ja,Lin:2008pr}. Further details on our ensembles can be found in Refs.\ \cite{Aarts:2014cda,Aarts:2014nba}. 
The main motivation for the next ensemble is to reduce the pion mass, keeping the other quantities unchanged (as much as possible). Following HadSpec \cite{Wilson:2015dqa,Cheung:2016bym}, we now use light quark masses which correspond to a pion of $m_\pi=236(2)$ MeV, with the strange quark mass unchanged. We refer to this as {\em Generation} 2L (L for light). A comparison of the lattice details is given in Table \ref{tab:1}. The important difference is the lighter pion and hence the larger extent in the spatial direction ($N_s=24\to 32$), to ensure a large enough physical volume. The renormalised anisotropies are nearly equal.

\begin{table}[h]
\centering
\begin{tabular}{l||cccccccc}
\hline
                & $a_s$ [fm]    & $a_\tau$ [fm]    & $a_\tau^{-1}$ [GeV]   & $\xi=a_s/a_\tau$  & $N_s$     &$m_\pi$ [MeV]  &   $m_\pi L$ \\
\hline
Generation 2    & 0.1227(8)     & 0.0350(2)  & 5.63(4)               & 3.5               & 24        & 384(4)    & 5.7 \\
Generation 2L   & 0.1136(6)     & 0.0330(2)  & 5.997(34)             & 3.453(6)          & 32        & 236(2)    & 4.3 \\
\hline
\end{tabular}
\caption{Comparison of lattice details for the Generation 2 and 2L ensembles.
} 
\label{tab:1}     
\end{table}

The Generation 2 ensembles were generated with Chroma \cite{Edwards:2004sx}. However, the lighter quarks proved a major stumbling block for Chroma. In order to generate finite-temperature ensembles, we have therefore adapted openQCD~\cite{openqcd}, which at the time had more advanced inversion algorithms, to include anisotropic lattices and stout-smeared gauge links. On top of this it makes use of additional AVX-512 optimisations, further improving runtime on recent Intel Skylake and Knights Landing CPUs~\cite{Bennett:2018oyb}. This adaptation of openQCD is publicly available\cite{fastsum,openqcd-fastsum}. In addition to this we made changes to openQCD so that it can be utilised as a framework for new lattice codes. We have used this to develop a stand-alone measurement code which constructs hadronic two-point functions \cite{openqcd-hadspec}. This software allows for the construction of correlation functions with and without Gaussian smearing at the sources and sinks, using the baryonic correlators defined in Ref.\ \cite{Leinweber}. The measurement code is available at the same location as our openQCD fork \cite{fastsum,openqcd-hadspec}.

The finite-temperature ensembles are listed in Table~\ref{tab:Gen2} and \ref{tab:Gen2L}. For Gen 2, the pseudo-critical temperature was determined via the renormalised Polyakov loop, and estimated to be $T_c=185(4)$ MeV. Hence there are four ensembles above and four below $T_c$. 
In the case of Gen 2L, the Polyakov loop no longer gives a clear location of the transition. Given that the light quarks are lighter, this is not unexpected. Below we will give first results for the transition as inferred from susceptibilities and the emergence of parity-doubling. In any case, for Gen 2L we have generated ensembles at 14 different temperatures, which will allow us to study the transition from the hadronic phase to the quark-gluon plasma in great detail.

\begin{table}[t]
\centering
\begin{tabular}{c||cccc | cccc}
\hline
 $N_\tau$ & 128$^*$ & 40 & 36 & 32 & 28 & 24 & 20 & 16 \\
\hline
$T$ [MeV]  		& 44 & 141  & 156   & 176 & 201  & 235  & 281 & 352 \\
$T/T_c$	        & 0.24 & 0.76  & 0.84   & 0.95 & 1.09  & 1.27  & 1.52 & 1.90 \\
$N_{\rm cfg}$ 	& 139  & 501 & 501 & 1000  & 1001 & 1001 & 1000 & 1001\\
\hline
\end{tabular}
\caption{Generation 2 ensembles, with lattice size $24^3 \times N_\tau$ \cite{Aarts:2014cda,Aarts:2014nba}. The ensemble at the lowest temperature has been provided by HadSpec \cite{Edwards:2008ja,Lin:2008pr}.}
\label{tab:Gen2}     
%\end{table}
\vspace*{0.4cm}
%\begin{table}[t]
\centering
\begin{tabular}{c||cccccccccccccc}
\hline
$N_\tau$        & 256$^*$ & 128 & 64 & 56 & 48 & 40 & 36 \\
\hline
$T$ [MeV]  		& 23 & 47 & 94 & 107 & 125 & 150 & 167 \\
$N_{\rm cfg}$ 	& 750 & 300 & 500 & 500 & 500 & 500 & 500 \\
\hline
$N_\tau$        & 32 & 28 & 24 & 20 & 16 & 12 & 8 \\
\hline
$T$ [MeV]  		& 187 & 214 & 250 & 300 & 375 & 500 & 750\\
$N_{\rm cfg}$ 	& 500 & 1000 & 1000 & 1000 & 1000 & 1000 & 1000  \\
\hline
\end{tabular}
\caption{Generation 2L ensembles, with lattice size $32^3 \times N_\tau$. The ensemble at the lowest temperature has been provided by HadSpec \cite{Wilson:2015dqa,Cheung:2016bym}.
} 
\label{tab:Gen2L}     
\end{table}

\section{Thermal transition: susceptibilities and parity-doubling}

\begin{figure}[t]
%    \vspace*{-1cm}
%    \begin{center}
%    \includegraphics[width=0.65\textwidth]{gfx/plot-chi-all-MeV-v2.pdf}
%    \vspace*{-0.3cm}
%    \caption{Preliminary results for the light and strange quark number susceptibilities (top) and charge and isospin %susceptibilities (bottom) on the Generation 2L ensembles, normalised with the corresponding expressions for massless %quarks on the lattice in the Stefan-Boltzmann limit \cite{Aarts:2014nba}.}
%    \label{fig:sus1}
%    \end{center}
%\end{figure}
\vspace*{-1cm}
%\begin{figure}[t]
    \begin{center}
    \includegraphics[width=0.8\textwidth]{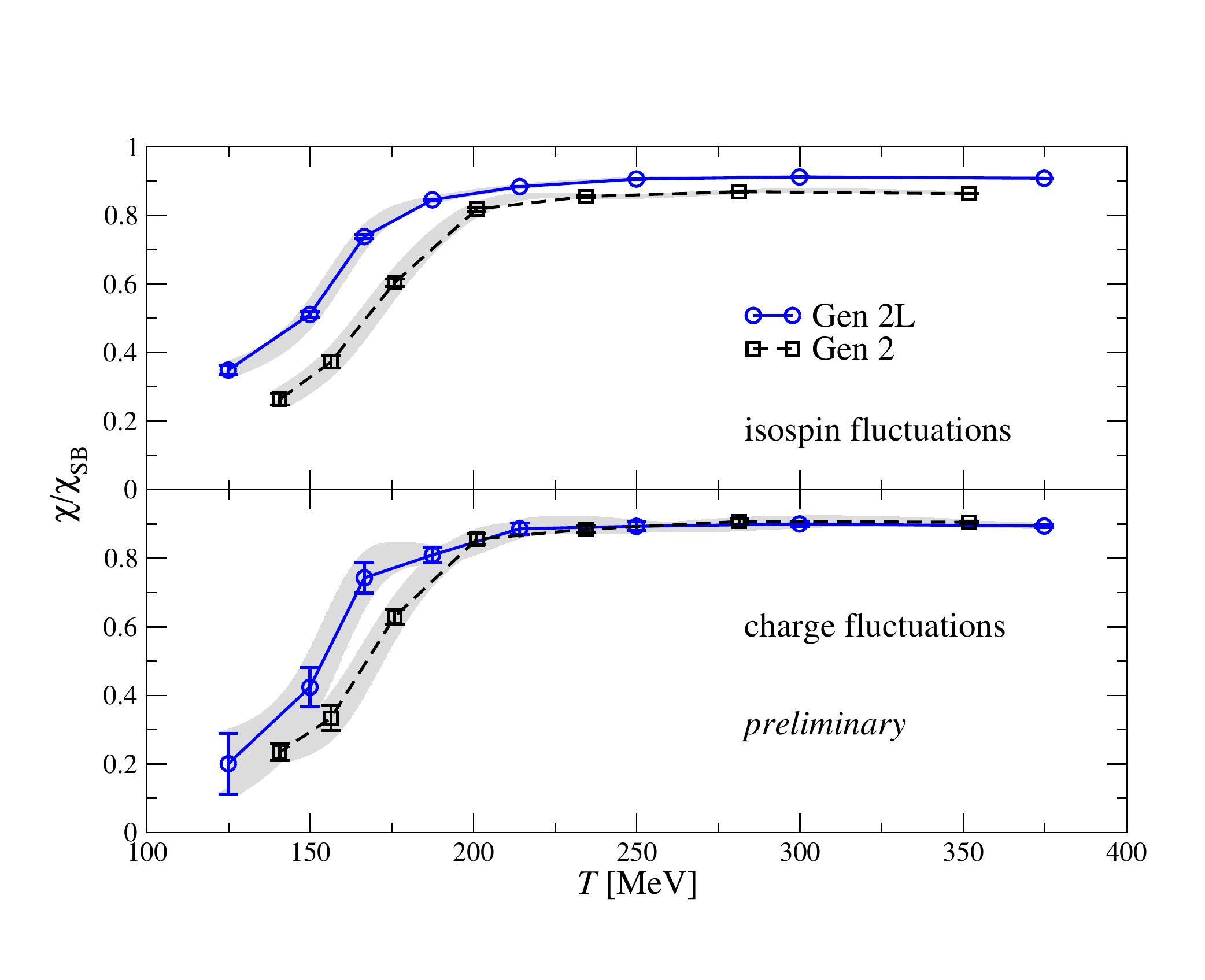}
    \vspace*{-0.3cm}
    \caption{Comparison of the isospin (top) and charge (bottom) susceptibilities on the Generation 2 and 2L ensembles. The shaded regions are cubic spline fits.
    }
    \label{fig:sus2}
    \end{center}
\end{figure}

To study the thermodynamic properties, we first discuss susceptibilities, i.e.\ fluctuations of light and strange quark number, and of baryon number, charge and isospin. We follow the approach described in Ref.\ \cite{Aarts:2014nba}. The computation is dominated by the stochastic estimates of disconnected contributions. Only for the isospin susceptibility, there is no such contribution and here the signal is cleanest. 
Preliminary results for the isospin and charge susceptibilities are given in Fig.\ \ref{fig:sus2}, where they are normalised with the corresponding quantities on the lattice for massless quarks in the Stefan-Boltzmann limit. The shaded regions are cubic spline fits.
The susceptibilities are qualitatively similar to those in Gen 2 \cite{Aarts:2014nba}, but some more effort is required to reduce the uncertainty. Note that not all temperatures are included at this stage. The main difference between Gen 2 and 2L is the shift of the transition region to lower temperatures, as illustrated in Fig.\ \ref{fig:sus2}, with the inflection points for both susceptibilities given by 
\be
T_{\rm infl} \simeq 155\; \mbox{MeV} \qquad \mbox{(Gen 2L)},
\qquad\qquad
T_{\rm infl} \simeq 169\;  \mbox{MeV} \qquad \mbox{(Gen 2)}.
\ee
Hence the reduction of the light quark masses brings this measure of the transition temperature closer to the pseudo-critical temperature observed with staggered fermions in the continuum limit \cite{Borsanyi:2010bp}. A study of the transition using twisted-mass Wilson fermions can be found in Ref.\ \cite{Burger:2018fvb}.
Isospin fluctuations are sensitive to the light quark masses, as observed in Fig.\ \ref{fig:sus2} (top). It is expected that the light quark mass dependence will eventually disappear, at higher temperatures.
More work is currently underway to include lower temperatures and reduce the uncertainty.

\begin{figure}[t]
    \vspace*{-1cm}
    \begin{center}
    \includegraphics[width=0.8\textwidth]{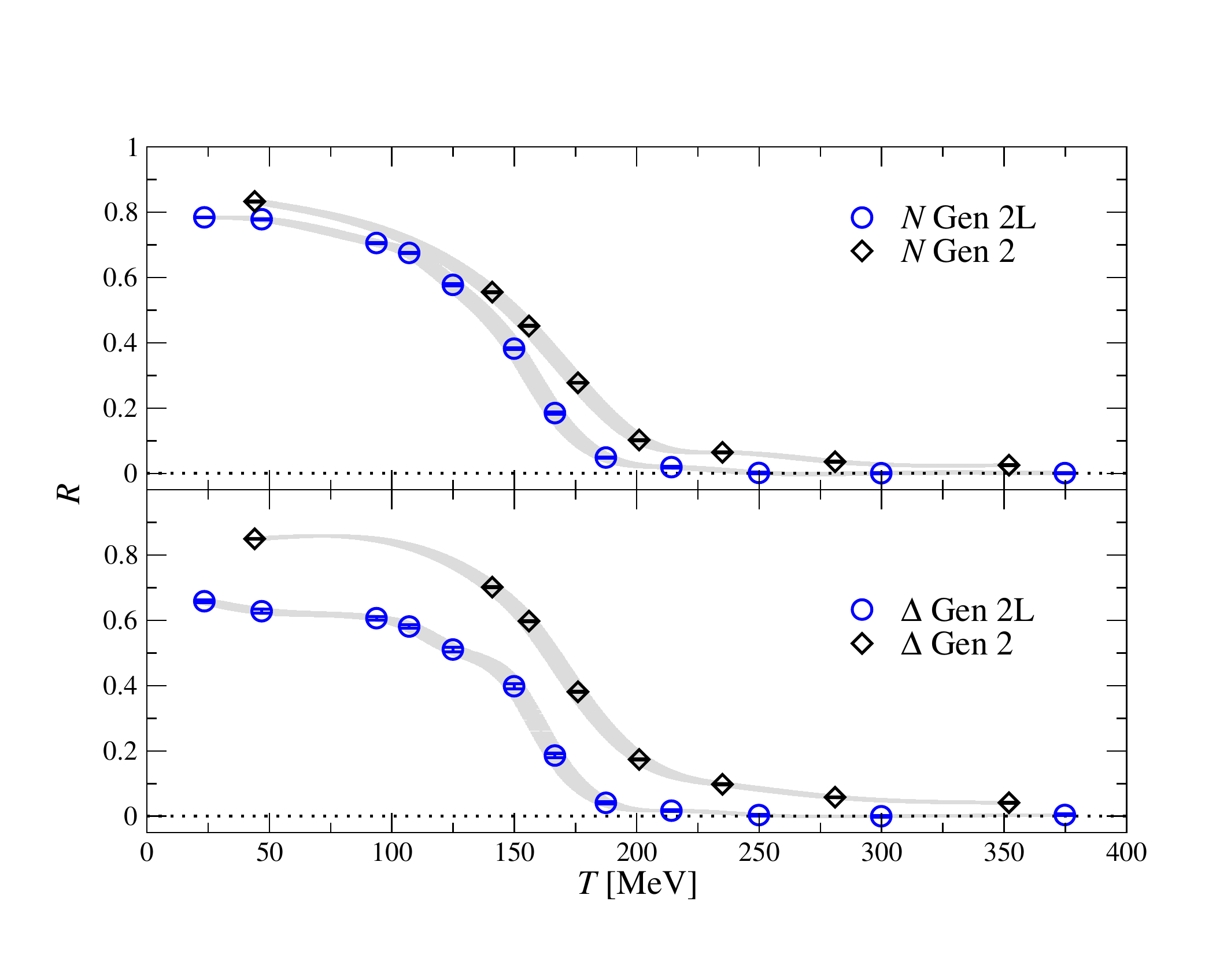}
    \vspace*{-0.3cm}
    \caption{Preliminary results for the comparison of the $R$ ratio in the nucleon (top) and $\Delta$ (bottom) channel, between the Generation 2 and 2L ensembles. The shaded regions are cubic spline fits.
    }
    \label{fig:Rcomp}
    \end{center}
\end{figure}

Another probe of the thermal transition is parity-doubling, which can be inferred directly from baryonic correlators, as discussed for the Gen 2 lattices in Refs.\ \cite{Aarts:2015mma,Aarts:2017rrl,Aarts:hype}. We construct the $R$ ratio from the positive- and negative-parity correlators $G_\pm(\tau)$ via \cite{Datta:2012fz,Aarts:2015mma}
\be
 R = \frac{\sum_n R(\tau_n)/\sigma^2(\tau_n)}{\sum_n 1/\sigma^2(\tau_n)},
\qquad\qquad\qquad
R(\tau) = \frac{G_+(\tau) - G_+(1/T-\tau)}{G_+(\tau) + G_+(1/T-\tau)},
\ee
where $G_+(1/T-\tau)=-G_-(\tau)$ and 
$\sigma(\tau_n)$ denotes the error at timeslice $\tau_n$. If chiral symmetry is unbroken, $R=0$. If chiral symmetry is broken and the mass of the negative-parity partner is substantially larger than the positive-parity one, $R\simeq 1$ \cite{Aarts:2015mma,Aarts:2017rrl,Aarts:hype}. 

A comparison between Gen 2 and 2L is shown in Fig.\ \ref{fig:Rcomp}, in the nucleon and the $\Delta$ channel. We observe a similar signal, but again with a shift of the transition to lower temperatures. The grey bands are obtained from cubic spline fits, which enables us to extract the temperature of the inflection point. In  both channels we find
\be
T_{\rm infl} \simeq 159\; \mbox{MeV} \qquad \mbox{(Gen 2L)},
\qquad\qquad
T_{\rm infl} \simeq 169\; \mbox{MeV} \qquad \mbox{(Gen 2)},
\ee
in line with the conclusions from the susceptibilities.
Finally, we note that in the Gen 2L ensembles $R\to 0$ at the higher temperatures more quickly than for Gen 2, which is again a manifestation of the quarks being lighter.

\section{Charmed baryons}

\begin{figure}[h]
    \vspace*{-1cm}
    \begin{center}
    \includegraphics[width=0.8\textwidth]{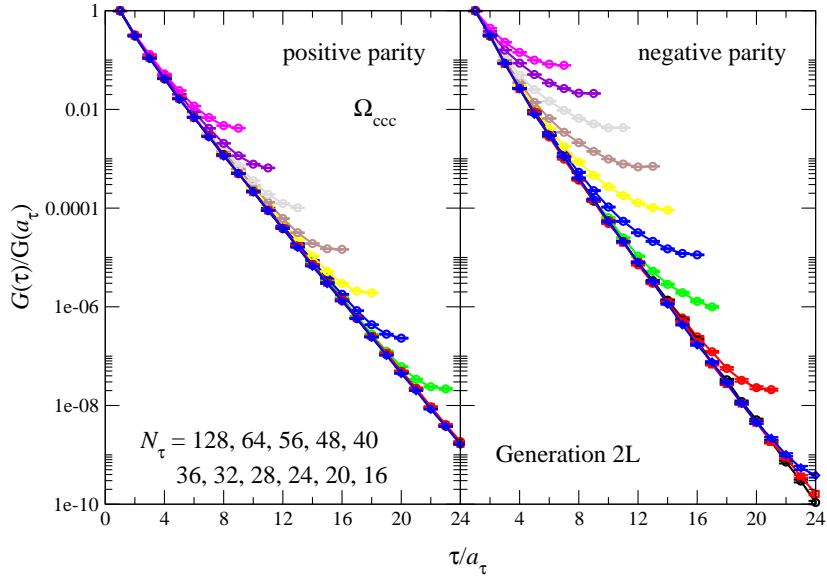}
    \vspace*{-0.3cm}
    \caption{Positive- and negative-parity correlators in the $\Omega_{\rm ccc}$ channel, at 11 different temperatures.}
        \vspace*{-0.5cm}
\label{fig:GOmega}
    \end{center}
\end{figure}

To probe further properties of the quark-gluon plasma, we have extended our previous studies of baryons to include charmed baryons (for previous studies at $T=0$, see e.g.\ Refs.\ \cite{Briceno:2012wt, Bali:2015lka}). One expects thermal effects to be less pronounced for the heavy charm quark, compared to light and strange quarks. Moreover, chiral symmetry remains explicitly broken. In Fig.\ \ref{fig:GOmega} we present first results for the positive- and negative-parity correlation functions of the $\Omega_{\rm ccc}$ baryon.
Interestingly, from the correlator itself one may already deduce that the groundstate in the positive-parity channel survives well into the quark-gluon plasma, whereas the negative-parity groundstate is more affected by the increase of temperature. A more quantitative study is currently underway.

\section{Summary}

We presented an update on our on-going projects on the FASTSUM ensembles of Generation 2L, with a lighter pion than previously considered, and showed preliminary results for susceptibilities and baryon correlation functions. A full study is on its way and will appear in due course.   

\section*{Acknowledgments}

We are grateful for support from STFC via grants  ST/L000369/1 and ST/P00055X/1, the Swansea Academy for Advanced Computing (SA$^2$C), SNF, ICHEC, and COST Action CA15213 THOR.
Computing resources were made available by HPC Wales and Supercomputing Wales and we acknowledge PRACE for access to the Marconi-KNL system hosted by CINECA, Italy.
This work used the DiRAC Extreme Scaling service and the DiRAC Blue Gene Q Shared Petaflop system at the University of Edinburgh, operated by the Edinburgh Parallel Computing Centre on behalf of the STFC DiRAC HPC Facility (www.dirac.ac.uk). This equipment was funded by by BIS National E-infrastructure capital grant ST/K000411/1, STFC capital grant ST/H008845/1, and STFC DiRAC Operations grants ST/K005804/1 and ST/K005790/1, and also through BEIS capital funding via STFC capital grant ST/R00238X/1 and STFC DiRAC Operations grant ST/R001006/1. DiRAC is part of the National e-Infrastructure. 

%\newpage
%% REMOVED to fit in on 7 pages! :) 

\end{document}